\documentclass{aims}

\usepackage{txfonts}
\usepackage{amsmath}
\usepackage{amssymb}
\usepackage{fontenc}
\usepackage{float}
\usepackage{graphicx}
\usepackage{amsfonts}
\usepackage{multicol}
\usepackage{xcolor}
\usepackage{color}
\usepackage{booktabs}
\usepackage{float}
\def\typeofarticle{Research article}
\def\currentvolume{x}
\def\currentissue{x}
\def\currentyear{x}

\def\ppages{xxx--xxx}
\def\DOI{}
\def\Received{}
\def\Revised{}
\def\Accepted{}
\def\Published{}

\numberwithin{equation}{section}

\begin{document}

\title{Stochastic volatility modeling of high-frequency CSI 300 index and dynamic jump prediction driven by machine learning}

\author{%
 Xianfei Hui\affil{1},
 Baiqing Sun\affil{1},
 Indranil SenGupta\affil{2},
 Yan Zhou\affil{1,}\corrauth
 and
 Hui Jiang\affil{3}
}

\shortauthors{Electronic Research Archive}

\address{%
 \addr{\affilnum{1}}{School of Management, Harbin Institute of Technology, Harbin 150001, China}
 \addr{\affilnum{2}}{ Department of Mathematics, North Dakota State University, Fargo, ND 58108-6050, USA}
  \addr{\affilnum{3}}{College of Management and Economics, Tianjin University, Tianjin 300072, China}
 }

\corraddr{zyhittxzz@126.com.}


\begin{abstract}
This paper models stochastic process of price time series of $CSI 300$ index in Chinese financial market, analyzes volatility characteristics of intraday high-frequency price data. {\color{black} In the new generalized Barndorff-Nielsen and Shephard model, the lag caused by asynchrony of market information and market microstructure noises are considered, and the problem of lack of long-term dependence is solved.} To speed up the valuation process, several machine learning and deep learning algorithms are used to estimate parameter and evaluate forecast results. Tracking historical jumps of different magnitudes offers promising avenues for simulating dynamic price processes and predicting future jumps. Numerical results show that the deterministic component of stochastic volatility processes would always be captured over short and longer-term windows. {\color{black}Research finding could be suitable for influence investors and regulators interested in predicting market dynamics based on high-frequency realized volatility. }
\end{abstract}


\maketitle

\section{Introduction}

As we all know, financial fluctuations may come not only from the financial system itself, but also from other aspects of social and economic life.
For example, COVID-19, has caused frequent and violent fluctuations in global financial markets \cite{Albulescu,Corbet}.
In the post-COVID-19 era, affected by internal and external factors in the market, the price of financial assets has been unstable during the first half of 2021.
Facing a world with more dynamic economic situation, enterprises and research circles are realising the importance of the challenges and opportunities presented by financial fluctuations.
The volatility of financial assets, which is the intensity of changes in the rate of return of financial assets over a period of time,
is unobservable \cite{Chauvet1}.
The measurement of volatility, which describes the potential deviation from the expected value, is the core issue in the study of financial volatility.
The accurate prediction of financial volatility is the key factor for successful financial asset pricing \cite{he2021closed}, economic forecasting \cite{Chauvet2}, risk management \cite{Arellano}, portfolio optimization \cite{Lin}, and quantitative investment \cite{caselli2020diversification}.
Volatility Analysis of financial time series is a practical method to study the law of volatility and estimate volatility.

{\color{black} An effective way to fit dynamic asset changes is stochastic volatility modeling in the research of financial quantification. There are a lot of derivative pricing models that could be used to model stock prices. He et al. (2021) \cite{he2023new} proposed a new stochastic volatility model to provide a better fit to real data and showed numerically the validity by comparing the results with the Monte Carlo simulation results. He et al. (2021) \cite{he2021fractional} used the FMLS (finite moment log-stable) model with the stochastic volatility to analyse the effect of both jumps and stochastic volatility. The numerical experiments that it is effective and converges very rapidly. Most interestingly, new progress has been made in the field of volatility estimation under high frequency environment in recent years.
A large amount of literature focuses on the three directions of high-frequency volatility estimation,
namely, model establishment \cite{he2022pricing}, model evaluation \cite{baker2021international}, and model application \cite{Barunik}.}
Asset price process \cite{Dutta} (continuous, finite jump or L\'evy jump) and data characteristics \cite{Jacod2}
(whether there is microstructure noise and whether regular sampling is implemented)
are the two main contents in the field of volatility estimation under high frequency environment.

Jump, excessive fluctuation of asset price in a certain period of time,
is one of the key issues in asset price dynamics research.
Theoretically, when there is no jump in asset price, the realized fluctuation is an unbiased and consistent estimation of potential fluctuation.
However, the jump phenomenon of price volatility in the capital market is widespread.
The jump leads to consistent overestimation of continuity fluctuations,
causes realized volatility and realized range volatility to no longer be an unbiased and consistent estimate of potential volatility.
In response to this jump phenomenon of asset price fluctuations,
estimation of realized bipower variation, which was first proposed by Barndorff-Nielsen and Shephard,
was used to decompose realized fluctuations into continuous fluctuations and jump fluctuations \cite{Barndorff}.

The Barndorff-Nielsen and Shephard model (BN-S) model \cite{Barndorff1},
which is used to describe the random behavior of price process in the research field of non-parametric methods for high-frequency time series,
is a popular stochastic volatility model with a L\'evy process as driving factor of financial asset price.
From academic points of view, the classic BN-S model has many attractive properties.
But its theoretical framework is not completely satisfied in many application scenarios.
Problems such as lack of long-term dependence may lead to the failure of the model in use.
Recently, a variety of improvement schemes to the basic model are proposed, generalized BN-S models are constructed,
and multiple dimensional applications,
such as jump capture \cite{Roberts1,Roberts2}, pricing \cite{SenGupta1,Sengupta2},
and risk management \cite{SenGupta3,Arai},
are implemented in the process of random fluctuations in asset prices.

Artificial intelligence in big data environment provides new tools for financial research and enriches the previous research on volatility estimation \cite{Mullainathan}.
Over the past few years, data processing classifiers based on machine learning and deep learning have always shown excellent performance in the field of financial prediction \cite{qian2022adaptive}.
Machine learning \cite{Henrique} could be used to deal with the nonlinear relationship and interaction between variables, which would effectively solve the collinearity problem between variables. The nonlinearity and randomness in financial time series could be captured by machine learning, so machine learning algorithms show bigger prediction ability than linear models. Monfared et al. (2014) \cite{monfared2014volatility} used machine learning algorithms to predict the volatility of the National Association of Securities Dealers Automated Quotations index in the United States. The results show that machine learning could improve the prediction ability. Compared with machine learning, deep learning \cite{Zheng} has stronger optimization capabilities and more advantages when dealing with big data sets. Recurrent neural network (RNN) is widely used because of its strong learning capability, stability and simulation ability for volatility prediction. Liu (2019) \cite{liu2019novel} deployed RNN for predicting the volatility of the S\&P 500 index, and found that RNN has stronger prediction ability than Generalized AutoRegressive Conditional Heteroskedasticity (GARCH). Fischer et al. (2018) \cite{fischer2018deep} used Long Short Term Memory(LSTM) to predict out-of-sample directional movements for the S\&P 500, and found that the performance of the stock portfolio constructed by this method is better than that of other linear models. The contribution of this paper is that the proposed approach through the use of machine learning algorithms enhances the forecasting ability of high-frequency $CSI 300$ index volatility, which is essential for dynamic jump prediction in hedging, and purposes of arbitrage, and that the new model solves the problem of the classical BN-S model and enhances the forecasting efficiency of high-frequency $CSI 300$ index volatility by fusing machine learning and deep learning algorithms with a generalized BN-S model.

In the research of asset volatility under random uncertain environment,
the classic BN-S model including a single OU process is often constructed in the previous literature.
However, the model will fail in application due to the lack of long-term dependence.
Some existing studies solve this problem by superimposing the OU process,
but the actual economic significance of different stochastic processes is less considered.
The generalized BN-S model is used to study the volatility of daily sampled commodity prices by many authors in the US and European markets.
There are fewer relevant studies on the Asia-Pacific market,
and fewer relevant studies using the minute sampling frequency.

As one of the fastest-growing markets in the world, the Asia-Pacific securities market has attracted more and more attention \cite{Zhou}.
The $CSI$ $300$ index covers most of the domestic market value of China (the largest economy in the Asia-Pacific region),
and reflects the market's mainstream investment returns and changes in the trader structure.
The use of samples with high sampling frequency in the day can retain more market information and discover more detailed fluctuation characteristics caused by the impact of various information on the market.
This research focuses on the price dynamics of the $CSI$ $300$ index with high sampling frequency,
and uses the generalized BN-S model to quantitatively analyze the volatility process of financial time series
to capture the deterministic component of the random process of price fluctuations.
The impact of overnight information \cite{Todorova} on the market is avoided in data preprocessing.
Samples with high sampling frequency in the day are used to retain market information to a greater extent
and discover more volatility characteristics caused by abnormal information shocks on the market.

Our approach to exploring stochastic process of high-frequency asset price dynamics has several advantages.
First, the certainty element $(\theta)$ in the new model helps us freely fit stock index prices and dynamic volatility in a correlated but different way. Because the superposition of L\' evy process is considered, it can solve the problem that the classical BN-S model does not have enough dependence for a long time. Many previous studies believe that future volatility is unobservable and completely random. It is noteworthy that, in \cite{sengupta2021refinements}, a generalized BN-S model with the superposition of L\' evy process extracted deterministic component in daily random volatility of crude oil out of low frequency price time series. Is there a deterministic component in random volatility that can improve the accuracy of volatility estimation? This paper studies the extension and application of the BN-S model in high-frequency financial time series field. The high-frequency data incorporate more market information than the daily price. Model description shows that the generalized new BN-S model is more superior in long-term volatility fitting and forecasting. Numerical results show that the deterministic component of stochastic volatility processes would always be captured over short and longer-term windows.

Second, the characteristics of information transmission and high-frequency data in the real financial market are fully considered in this paper. The new model realize the estimation of delay parameter $(b)$ in the case of the jump in volatility caused by sluggish market response is not synchronized with the jump of asset price. Unlike low-frequency random volatility models, high-frequency random volatility models need to consider the impact of market microstructure noises based on data characteristics. Overnight information is another key factor affecting the accuracy of prediction results in high-frequency data processing. The impacts of market microstructure noises and overnight information are also considered in data analysis. The numerical results on $CSI$ $300$ index show the effectiveness of the generalized BN-S model when it is used for high-frequency data with noise.

Third, the new model supports the cooperation of machine learning and stochastic volatility models, applies big data techniques to feature learning and parameter estimation on empirical datasets of stock index prices, and captures the deterministic components of the intraday price volatility of $CSI$ $300$ stock index. It is easy to estimate the dynamic deterministic parameter with the help of machine learning algorithms and deep learning algorithms. It shows the application of data science in obtaining ``deterministic components" from processes that are generally considered to be completely random. 11 kinds of machine learning and deep learning algorithms (Logistic regression, Support vector machine, K-nearest neighbors, K-means, Naive bayes, Gradient boost, Decision tree, Random forest, Neural network, LSTM and LSTM network with batch) are used to process data to estimate parameters. We believe that more algorithms could further prove the validity of the model and ensure the accuracy of the results. In general, the results offer promising avenues for simulating dynamic price processes and predicting future jumps. Research finding could be suitable for influence investors and regulators interested in predicting market dynamics based on high-frequency realized volatility.

The paper is organized as follows. In Section \ref{sec2}, the generalized BN-S model is introduced.
In Section \ref{sec3}, the high-frequency $CSI$ $300$ stock index price data is selected as the research sample,
the high-frequency financial time series are preprocessed,
the descriptive statistical characteristics of the data set are obtained,
and the distribution of price fluctuations is analysed.
Based on the research results obtained in Section \ref{sec3},
a deterministic component out of high-frequency price stochastic processes is derived
by using machine learning and deep learning algorithms to realize parameter analysis and estimation in Section \ref{sec4}.
In Section \ref{sec5}, a brief conclusion is provided.

\section{Barndorff-Nielsen and Shephard model}
\label{sec2}
Financial time series of different assets share many common features (heavy tailed distributions of log-returns, aggregational gaussianity, quasi long-range dependence). Many of these facts are successfully captured by stochastic models with L\'evy processes.
L\'evy processes can be used to characterize the dynamic changes of the time series of financial asset prices with jump processes.
BN-S model, which is a widely used stochastic model with L\'evy processes,
is used to describe the stochastic behavior of random time series in the research field of nonparametric methods of high-frequency time series.
A brief introduction to this model is given as follows.

Consider a frictionless financial market in which a risk-free asset with a constant rate of return $r$ and a stock are traded on a fixed horizon date $T$.
The classical BN-S model assumes that the price process of a stock (or, a commodity) $S=(S_t)_{t \ge 0}$, which is defined in a filtered probability space
$(\Omega, \mathcal{F}, (\mathcal{F}_t)_{0\le t \le T}, \mathbb{P})$, is given by
\begin{equation}
S_t=S_0 \exp(X_t),
\label{(2.1)}
\end{equation}
the log-return $X_t$ is given by
\begin{equation}
dX_t=(\mu + \beta \sigma_t^2)dt+\sigma_t d W_t+\rho d Z_{\lambda t},
\label{(2.2)}
\end{equation}
where $\sigma_t$ is the volatility at time $t$, the parameters $\mu, \beta, \rho\in \mathbb{R}$, and $\rho \le 0$.
The variance process is given by
\begin{equation}
d\sigma_t^2=-\lambda \sigma_t^2 dt+dZ_{\lambda t}, \quad \sigma_0^2>0,
\label{(2.3)}
\end{equation}
where $\lambda> 0$.

With respect to the probability measure $ \mathbb{P}$, the process $W = (W_t)$ is a standard Brownian motion.
Observe that the Ornstein-Uhlenbeck process in this model is driven by an incremental L\'evy process,
which is a random process of positive mean recovery.
The process $Z=(Z_{\lambda t})$ is the subordinator (also known as \emph{background driving L\'evy process} or ``BDLP"). The processes $W$ and $Z$ are independent of each other. Also, $({\cal F}_t)$ is a conventional augmentation of the filtering produced by $(W, Z)$.

Solving \eqref{(2.3)} we obtain
\begin{equation}
 \sigma_t^2=e^{-\lambda t}+\int_{0}^{t} e^{-\lambda (t-s)}\, dZ_{\lambda s}.
\label{(2.4)}
\end{equation}
Clearly, the process $\sigma^2=(\sigma_t^2)$ is strictly positive. The classical BN-S model has excellent performance in describing the dynamic characteristic response mode of stable asset prices in a short time.
It is commonly used to capture some stylized features of time series observed in financial markets, such as semiheavy tails, aggregational Gaussianity, quasi long range dependency and self-similarity.

However, the results and theoretical framework of the classical BN-S model are not completely satisfactory in empirical situations. There are several problems in the classical model, which may make the model difficult to use in practice. For example, empirical results show that the jump in volatility is positively correlated with stock or commodity prices.
But the jump phenomenon in volatility does not usually occur at the same time with the change in price because of the lag in market response.
The topic of delayed response in the financial market has been studied in papers such as \cite{Grobys}. On the other hand, the study in \cite{xiao2020dynamic} handles this problem with a delayed price formula, where the price volatility obeys the form $\sigma(S_t-b)$, for some delay parameter $b > 0$. However, the parameter $b$ is also stochastic, and this makes the resulting model unnecessarily involved.

Furthermore, the classical BN-S model does not have long-term dependence property. Consequently, due to the high sequence correlation between hidden variables and parameters, for the analysis of the empirical data based on this model the convergence rate is slow. The classical BN-S model contains a single BDLP, which makes the logarithmic return, volatility and variance in the model completely dependent on each other. When the model is used over a long period of time, this absolute correlation may lead to inaccurate results. As a result, the model encounters serious failures in volatility estimation.

These problems are overcome in a new generalized model. It is clear that for the long-term implementation of the classical BN-S model, a single L\'evy subordination is obviously ineffective.
The research results \cite{Habtemicae1} show that the superposition of Ornstein-Uhlenbeck (OU) type processes can achieve long-range dependence.
The superposition of L\'evy subordinations successfully fits the asynchronous changes from price and volatility in an interrelated but independent way. Referencing the previous research results \cite{sengupta2021refinements}, the structure of a generalized BN-S model will be introduced as follows.

The key point of our research is to capture the deterministic components out of high-frequency price stochastic processes.
As proposed in \cite{sengupta2021refinements}, suppose $Z_t$ and $Z_t^ {*}$ , with same (finite) variance, are two independent L\'evy subordinators.
There exists a L\'evy subordinate $\overline{Z}_{\lambda t}$ independent of $W$, such that
\begin{equation}
 d\overline{Z}_{\lambda t}= \rho' d Z_{\lambda t}+\sqrt{1-\rho'^2} d Z_{\lambda t}^{*} , \quad 0\le \rho' \le 1.
\label{(2.5)}
\end{equation}
For $0 \le \rho' \le 1$, $Z$ and $\overline{Z}$ are positively correlated L\'evy subordinators.
Assume that the dynamics of $S_t$ are given by \eqref{(2.1)} and \eqref{(2.2)}, where $\sigma_t$ is given by
\begin{equation}
 d\sigma_t^2= -\lambda \sigma_t^2 dt+ d\overline{Z}_{\lambda t}, \quad \sigma_0^2 >0.
\label{(2.6)}
\end{equation}
In \eqref{(2.6)}, the OU process $\overline{Z}=(\overline{Z}_{\lambda t})$ is related to the corresponding $Z$ in \eqref{(2.3)} and is also independent of $W$.

In the following study, delay parameter b and the long range dependence property of model are considered. As shown in \cite{sengupta2021refinements}, the price $S=(S_t)_{t\ge 0}$ on some risk-neutral filtered probability space $(\Omega, \mathcal{F}, (\mathcal{F}_t)_{0\le t \le T}, \mathbb{P})$ is modeled by \eqref{(2.1)}. And the convex combination of two independent subordinators $Z$ and $ Z^ {(b)}$ would be implemented to expressed the dynamics of $X_t$ in (2.2) by
\begin{equation}
 dX_t= (\mu +\beta \sigma_t^2)dt +\sigma_t d W_t +\rho ((1-\theta)dZ_{\lambda t}+\theta d Z_{\lambda t}^{(b)}),
\label{(2.7)}
\end{equation}
where $0 \le \theta \le 1$, $\theta$ is a deterministic parameter.
At time $t$, $\lambda >0$ is the proportional parameter. $Z_{\lambda t}$ and $Z_{\lambda t}^{(b)}$ are independent L\'evy processes.
Compared to $Z_{\lambda t}$, the process $Z_{\lambda t}^{(b)}$ corresponds to the greater L\'evy intensity.
For instance, if the L\'evy densities of $Z$ and $Z^{(b)}$ are given by $v_1ae^{-ax}$ and $ v_2ae^{-ax}$, respectively
(for $a> 0,\ v_1> 0,\ v_2 > 0,$ and $x > 0$),
then $v_2 > v_1$. Also, in \eqref{(2.8)} the processes $W, Z$ and $Z^{(b)}$are independent, and $({\cal F}_t)$ is the usual augmentation of the filtration generated by $(W;Z; Z^{(b)})$.

{\color{black} In fact, the real financial market often deviates from the efficient market. The short-term changes of asset prices are usually affected by various market factors (e.g. transaction cost, asymmetric information of traders, etc.), which makes the observed prices deviate from the real prices. This deviation is called market microstructure noise. The observed asset price process, generally bought in noise, could be divided into two components }
\begin{equation}
\overline{X}_t=X_t+\varepsilon_t,
\label{(2.8)}
\end{equation}
{\color{black}where $\overline{X}_t$ is the observed log price, $X_t$ is the factual true value of log price, and $\varepsilon_t$ is market microstructure noise. The asset log price processes and market microstructure noise are independent of each other. $\varepsilon_t$ is independent identically distributed, $E \varepsilon_t=0$ and $Var(\varepsilon_t)=E \varepsilon_t^2$. Market microstructure noise has a significant impact on estimation of high-frequency covariance matrix. Market microstructure noise has a significant impact on estimation of high-frequency covariance matrix. For high sampling frequency, the estimated value of realized covariance is in fact not the covariance of assets price, but the covariance of market microstructure noise. The asset price process $S_t$ in \eqref{(2.1)} is given by}

\begin{equation}
S_t=S_0 exp(\overline{X}_t-\varepsilon_t).
\label{(2.9)}
\end{equation}
{\color{black}The convex combination of two independent subordinators $Z_{\lambda t}$ and $Z_{\lambda t}^{(b)}$ would be implemented to expressed the dynamics of log price in \eqref{(2.7)} is given by}

\begin{equation}
d(\overline{X}_t-\varepsilon)=(\mu +\beta \sigma_t^2)dt+\sigma_t d W_t +\rho((1-\theta)dZ_{\lambda t}+\theta d Z_{\lambda t }^{(b)}).
\label{(2.10)}
\end{equation}

The variance process in \eqref{(2.3)} in this case is given by
\begin{equation}
d\sigma_t^2= -\lambda \sigma_t^2 dt+ (1-\theta')dZ_{\lambda t}+\theta^{'} d Z_{\lambda t}^{(b)},\sigma_0^2 >0,
\label{(2.11)}
\end{equation}
where $\theta' \in [0,1]$ is deterministic parameter. For simplicity, assume $\theta=\theta^{'}$ for the rest of this paper. The sum of $(1-\theta')dZ_{\lambda t}$ and $\theta' d Z_{\lambda t}^{(b)}$, is a L\'evy process, which is positively correlated with $Z_{\lambda t}$
and $Z_{\lambda t}^{(b)}$.

After a simple calculation, the solution of \eqref{(2.11)} can be explicitly written as
\begin{equation}
\sigma_t^2=e^{-\lambda t}\sigma_0^2+\int_{0}^{t} e^{-\lambda (t-s)}\, ((1-\theta^{'})dZ_{\lambda t}+\theta^{'} d Z_{\lambda t}^{(b)}).
\label{(2.12)}
\end{equation}
This enforces positivity of $\sigma_t^2$.
Thus, the process $\sigma_t^2$ is strictly positive and it is bounded from below by the deterministic function $e^{-\lambda t}\sigma_0^2$.
The instantaneous variance of log returns is given by
$$
(\sigma_t^2+\rho^2(1-\theta)^2 \lambda Var[Z_1]+\rho^2 \theta^2 \lambda Var[Z_1^{(b)}])dt.
$$

The short-range-dependence problem of the classical BN-S model can be improved in the new model. The dynamics given by the new model incorporates a long-range dependence. Assume that $J_Z$ is a jump measure related to the subordinate $Z$ of the L\'evy process,
$J_Z^{(b)}$ corresponds to the subordinate $Z^{(b)}$ of the L\'evy process,
and $J(s)=\int_0^s \int_{\mathbb{R}+}J_Z(\lambda d\tau ,dy)$,
$J^{(b)}_{(s)}=\int_0^s \int_{\mathbb{R}+}J_Z^{(b)}(\lambda d\tau ,dy)$.
Considering the logarithmic regression of the classical BN-S model and the new model, the covariances of $X_t$ and $X_s$ are given by

\begin{equation}
Corr(X_t,X_s)= \frac{\int_0^s \sigma_\tau^2d\tau +\rho^2J(s)}{\sqrt{(\int_0^t \sigma_\tau^2 d \tau + t \rho^2 \lambda Var(Z_1))(\int_0^s \sigma_\tau^2 d \tau + s \rho^2 \lambda Var(Z_1))}}, \quad t>s
\label{(2.13)}
\end{equation}
and
\begin{equation}
Corr(X_t,X_s)= \frac{\int_0^s \sigma_\tau^2d\tau +\rho^2(1-\theta)^2J(s)+\rho^2 \theta^2 J^{(b)}(s)}{\sqrt{\alpha(t)\alpha(s)}}, \quad t>s
\label{(2.14)}
\end{equation}
respectively, where $\alpha(\nu)=\int_0^\nu \sigma^2_\tau d\tau +\nu \rho^2 \lambda((1-\theta)^2Var(Z_1)+\theta^2Var(Z_1^{(b)})).$
When $s$ takes a fixed value, for the classical BN-S model, $Corr(X_t, X_s)$ rapidly becomes smaller with the increase of $t$. Such attenuation may cause the failure of the classical model in applications with a long time span. It can be seen that the BN-S model, when used to fit the random fluctuation process of risky assets, may get inaccurate fluctuation simulation results, affected by the change of the time parameter $t$.

On the other hand, variance of the log-returns $X_t$ and $X_s$ (as shown in \eqref{(2.14)}) are
$$
\int_{0}^{t} \sigma_{\tau}^2 d \tau \, + \nu \rho^2 \lambda ((1-\theta)^2 Var (Z_1) + \theta^2 Var (Z_1^{(b)}))
 $$
and
$$
\int_{0}^{s} \sigma_{\tau}^2 d \tau \, + \nu \rho^2 \lambda ((1-\theta)^2 Var (Z_1) + \theta^2 Var (Z_1^{(b)}))
 $$
respectively.

Affected by the value of the parameter $\theta$, $Corr (X_t, X_s)$ will never become ``too small". Because the value of $t$ must have an upper limit when $s$ takes a fixed value.
This is the main difference between the results of \eqref{(2.13)} and \eqref{(2.14)}.
It can be clearly seen from the results that the generalized new BN-S model incorporates a long-range dependence and provides more accurate characteristics for the dynamic volatility analysis in the asset price process. The new model can accurately capture the essential characteristics of the random fluctuation process of financial time series.

In addition, compared to the classical BN-S model, the parameter $\theta$ in the new model can help us freely fit asset prices and volatility in a correlated but different way.
For dynamic prices, the jump is not completely random, and there is a deterministic element $(\theta)$
that can be implemented to be effectively applied to the new BN-S model in a longer time.
The large fluctuations can be captured in the future from historical experience data $(\theta=1)$,
and the initial L\'evy subordinate function $Z_{\lambda t}$ could be converted into a stronger L\'evy subordinate function $Z_{\lambda t}^{(b)}$
to correspond to the large fluctuations.
If there is no big jump apprehended for the upcoming time, the L\'evy subordinate function $Z_{\lambda t}^{(b)}$ could be converted into L\'evy subordinate function $Z_{\lambda t}$ based on historical data $(\theta=0)$ by using machine learning and deep learning algorithms.

Obviously, an important challenge in the application of the new model is to obtain an estimate of the value of a deterministic component of the empirical data.
In this paper, the new model is used to analyze the price dynamics of high frequency $CSI$ $300$ stock index.
Several machine learning algorithms and deep learning algorithms are implemented to forecast parameter $\theta$.

\section{Data}
\label{sec3}

\subsection{Sources of data}

The $CSI$ $300$ index is always considered to have strong market representation.
It covers most of China's domestic circulating market value and reflects the overall trend of China's Shanghai and Shenzhen markets.
In particular, its constituent stocks include many mainstream investment stocks with market representation, liquidity and trading activity.
So it is often used to study the returns of mainstream investments and changes in financial price fluctuations in the market.

\begin{figure}[h!]
\centering
{\includegraphics[width=6in,height=3in]{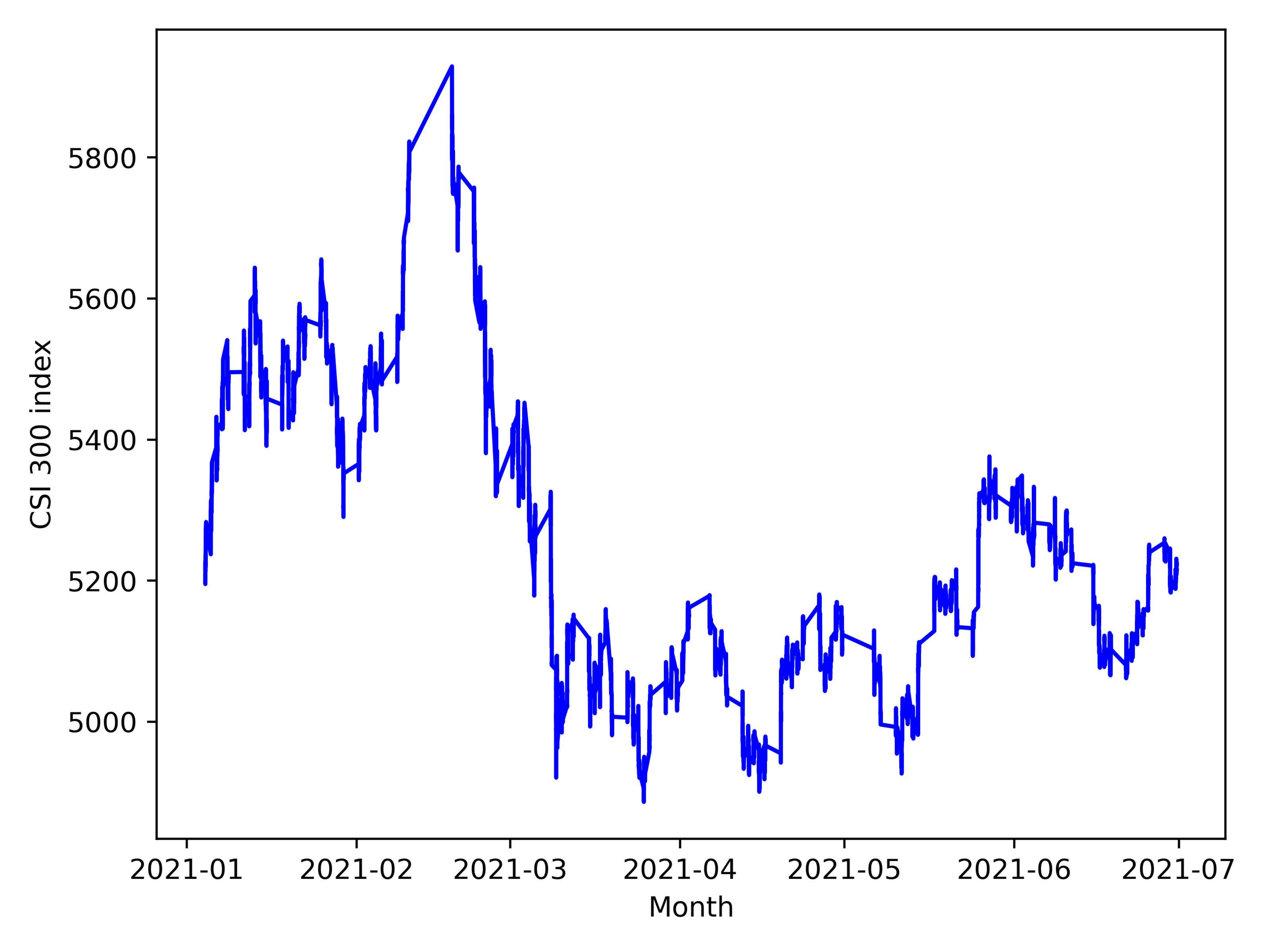}}
\caption{ \small{Curve of closing price per minute}}
\end{figure}

The main purpose of this paper is to explore the volatility characteristics of intra-day high-frequency price data, and then to study the quantitative indicators in the process of random fluctuations in financial time series. The generality and extensiveness of the application of the new model in the previous section are considered, and the $CSI$ $300$ index price is considered as the empirical data of analysis. The corresponding intra-day high-frequency data is selected as the research sample.
It is conducive to maximally retain market information to select research samples with a higher sampling frequency.
The intraday closing price data of the $CSI$ $300$ index on consecutive trading days from January $1$, $2021$ to June $30$, $2021$ is considered as a sample.
The sampling frequency of this sample is $1$ minute.
The data set contains a total of $28,320 $ observations in $118$ consecutive trading days (Data source: Wind\footnote{In the field of financial data, Wind has built a complete and accurate large-scale financial engineering and financial data center on financial and securities data in China. Uniform Resource Locator of wind is: https://www.wind.com}).

The fluctuation curve of the historical data over time is shown in Figure 1.

It is necessary to discuss the data distribution characteristics of intra-day price changes and yield fluctuations, which help us to explore the basic laws of $CSI$ $300$ stock index time series fluctuations. In order to study the change trend and distribution characteristics of $CSI$ $300$ stock index over time in different time intervals, an intuitive way is chosen to visually analyze the data structure. Figure 2 shows the moving average curve of the $CSI$ $300$ index under different time spans.

Normally, the trading hours of the $CSI$ $300$ index are each working day in $9:30-11:30$ and $13:00-15:00$, Beijing time (the effective trading time per day is a total of $4$ hours).
So four time spans of $1$ minute, $30$ minutes, $120$ minutes (half a day) and $240$ minutes ($1$ day) are chosen to observe the data set.
In Figure 2, blue represents the price change curve per minute, and red represents daily price fluctuations.
It can be clearly seen that the general trends of the two curves are similar, but there are fewer repetitions.
The blue curve fluctuates more sharply than the red one,
which shows that the high-frequency data during the day contains more market information than the closing price.
The yellow line (representing the price change every $30$ minutes) and the blue line overlap more severely than the red line.
The green curve, which represents price changes every $2$ hours, is more stable than the yellow line.
These results are also considered to confirm the above view,
that is, the data set at a higher sampling frequency is more effective for us to find the realized volatility estimator.
Compared with previous studies, the data set used in this paper has more advantages.

\begin{figure}[h!]
\centering
{\includegraphics[width=6in,height=3in]{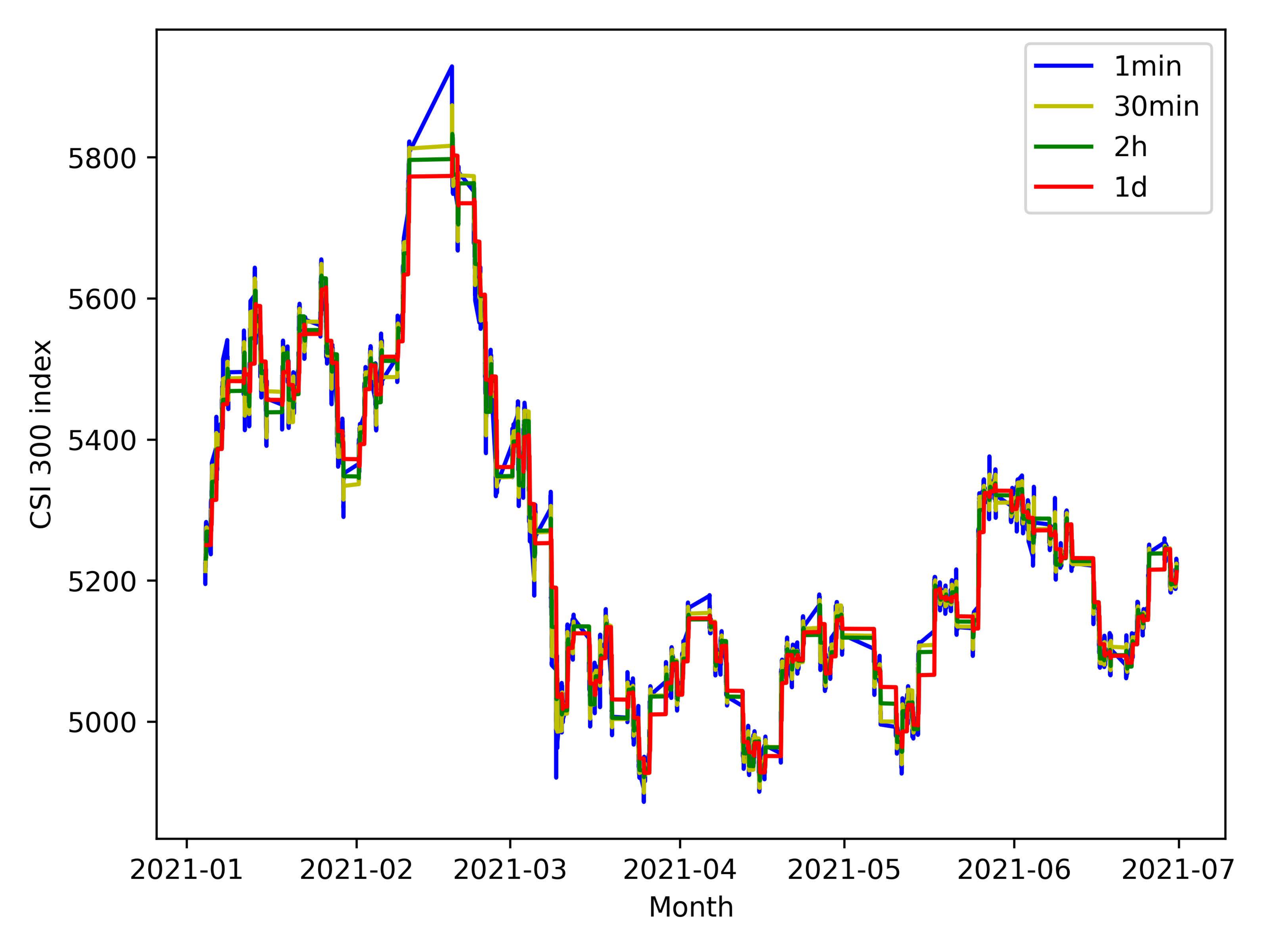}}
\caption{ \small{Moving average for $CSI$ $300$ index}}
\end{figure}

\subsection{Data preprocessing}

A lot of misleading information exists in the unprocessed empirical data for various reasons.
Unprocessed data is used directly, which may lead to undesirable results such as a decrease in the prediction accuracy of the time series.
Therefore, the observed samples should be filtered before doing data analysis.

Compared with the fluctuations in daily stock index yields and trading volume,
the impact of overnight information on the market should not be ignored.
Most of these price changes on overnight information are concentrated within ten minutes of the opening.
In other words, price fluctuations within ten minutes of the opening could not represent changes in stock index fluctuations throughout the trading day.
In order to avoid shocking intra-day fluctuations and causing abnormal data (such as high kurtosis, increasing outliers, etc.),
the data within $10$ minutes of the opening of the daily observation sample should be excluded.
After the overnight information was digested by the market, the empirical data would reflect the daily operation of the $CSI$ $300$ index price more accurately. In addition, we remove the outliers and zero-value data from the observed data to keep the observed sample data tidy.

After preprocessing the empirical data according to the above filter conditions, the usable sample data are filtered out ($28081$ observations in total). The rejection rate of sample data is $0.84\%$. It shows that the observed samples have both liquidity and validity, and the intra-day high-frequency price information is effectively stored in the empirical data.

\subsection{Descriptive analysis}

For the convenience of research, the observed samples are divided and numbered in chronological order. For example, Sample 1, Sample 2, $\dots$, Sample 6, represent the samples from January through June of $2021$. {\color{black}Variance of realized volatility and realized covariance generally decrease with increased sampling frequency. Market microstructure noise has a significant impact on estimation of high-frequency covariance. Therefore, in the realized covariance estimation of financial high-frequency data, it is necessary to balance the variance and bias to select the appropriate sampling frequency. As proposed in \cite{andersen2003modeling}, the sampling frequency of 5 minutes is selected in this paper.} The statistical descriptions of the samples of high-frequency $CSI$ $300$ index intraday prices are given in Table 1.

\begin{table}[H]
\center
\color{black}
\caption{Statistical description of $CSI$ $300$ index high-frequency prices}
\resizebox{\textwidth}{30mm}{
\begin{tabular}{cccccccc}
 \toprule
 & Overall sample & Sample 1 	& Sample 2 & Sample 3 & Sample 4 & Sample 5 & Sample 6\\
 \midrule
Count & 28081& 940 &705 &1081 & 987 & 846 & 987\\
Mean & 5245.46 & 5471.63 & 5563.75 & 5116.65 & 5067.71 & 5151.67 & 5202.71  \\
Median  & 5187.84 & 5485.11 & 5523.59 & 5074.30 & 5087.45 & 5161.70 & 5221.97 \\
Minimum & 4886.40 & 5209.49 & 5319.68 & 4891.91 & 4902.15 & 4929.86 & 5065.91 \\
Maximum & 5908.34 & 5653.55 & 5894.03 & 5454.05 & 5177.60 & 5364.37 & 5343.28 \\
Skewness & 0.60 & -0.68 & 0.40 & 0.94 & -0.73 & 0.11 & -0.21 \\
Kurtosis  & -0.50 & 0.23 & -0.86 & -0.17 & -0.66 & -1.19 & -1.16 \\
\bottomrule
\end{tabular}}
\end{table}

The fluctuation characteristics of $CSI$ $300$ prices could be seen from the statistical results in Table 1.
The highest price was $5908.34$ in February, and the lowest price of $4886.4$ appeared in March.
Sample 2 has the least amount of data, but its mean and median are higher than those of other samples.
It shows that the price in February is more advantageous compared with other months.
The distribution of price data is not completely symmetrical.
The skewness of the overall sample of Sample 1, Sample 4, and Sample 6, are all less than zero.
Their distributions have negative deviations, and the tail on the left is longer than the right.
Because there are a few variables with small values, the left tail of the curve is dragged longer.
In contrast to them, there are heavy-tailed distributions on the right side of Sample 2, Sample 3, and Sample 5 (the skewness of these three samples are all greater than $0$).
This phenomenon is most obvious in May, followed by June.
The kurtosis of the observed samples are less than $3$, which shows that the observed samples do not have leptokurtic characteristics.
We believe this is related to sampling frequency.
In the case of sampling frequency per minute, the kurtosis of the sample is less than that of normal distribution.
The generalized BN-S model mentioned in Section 2 is suitable to discuss the above data characteristics,
because L\'evy processes in the model could be used to characterize the dynamic changes of the time series of financial asset prices with jump processes.

\begin{figure}[h!]
\centering
{\includegraphics[width=5.5in,height=3in]{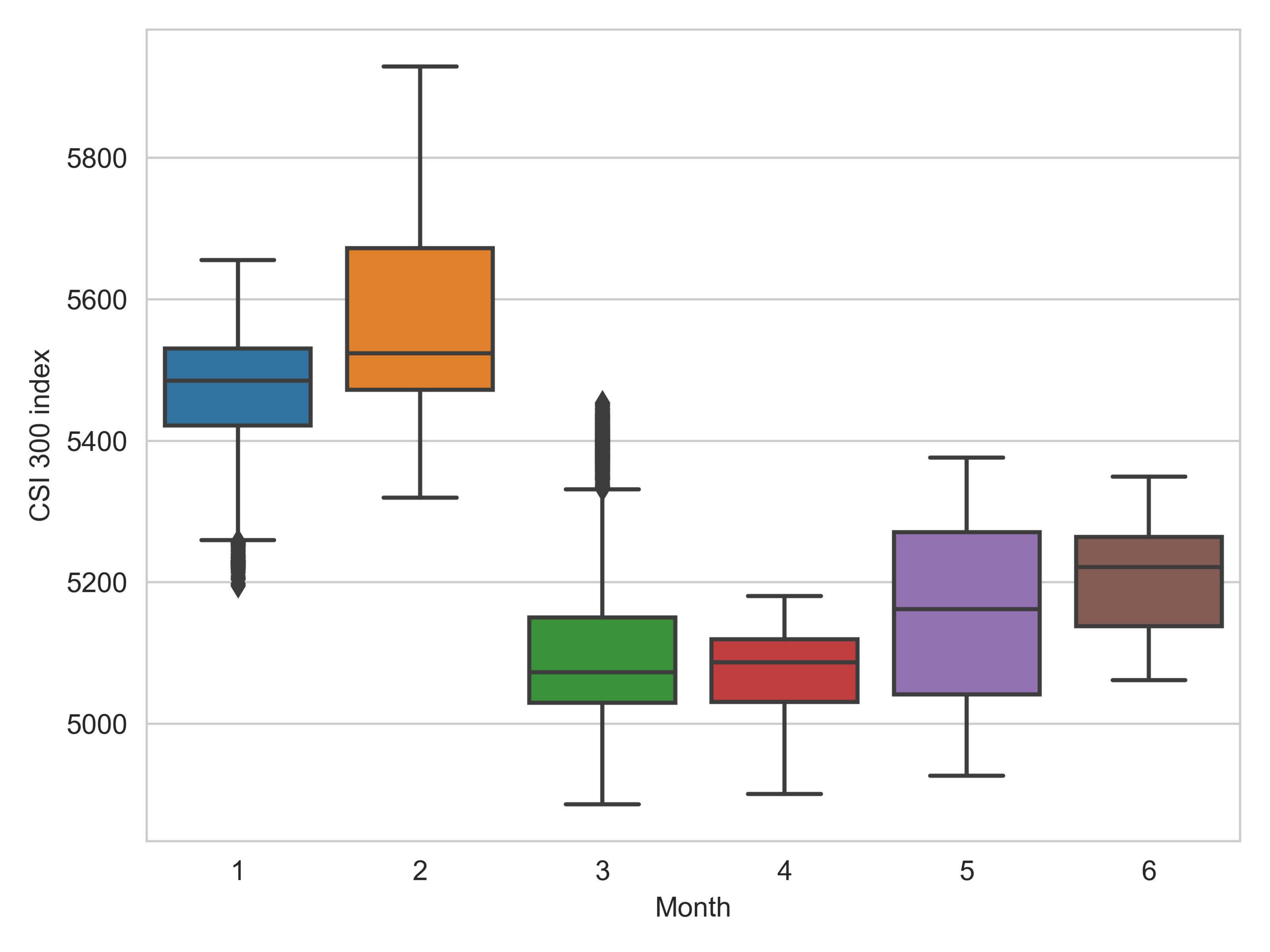}}
\caption{ \small{ Daily boxplot for $CSI$ $300$ index}}
\end{figure}

Figure 3 provides the difference in the distribution of intra-day high-frequency price samples of $CSI$ $300$ in different time periods through a box plot.
Compared with other samples, the prices in January and February are more advantageous,
and the price fluctuation in February is also the largest.

A histogram of price distribution explains the dispersion and distribution of $CSI$ $300$ in half a year in Figure 4.
Obviously, the $CSI$ $300$ index is the densest in the range of $5000-5200$.
Together with Figure 3, it could be seen that the prices from March to June are mostly within this range.
It shows that the price fluctuations from January to February are likely to be more volatile, and the fluctuates in the smooth from March to June.
The larger jumps we are concerned about are most likely to occur in Sample 1 and Sample 2.

\begin{figure}[h!]
\centering
{\includegraphics[width=5.5in,height=3.5in]{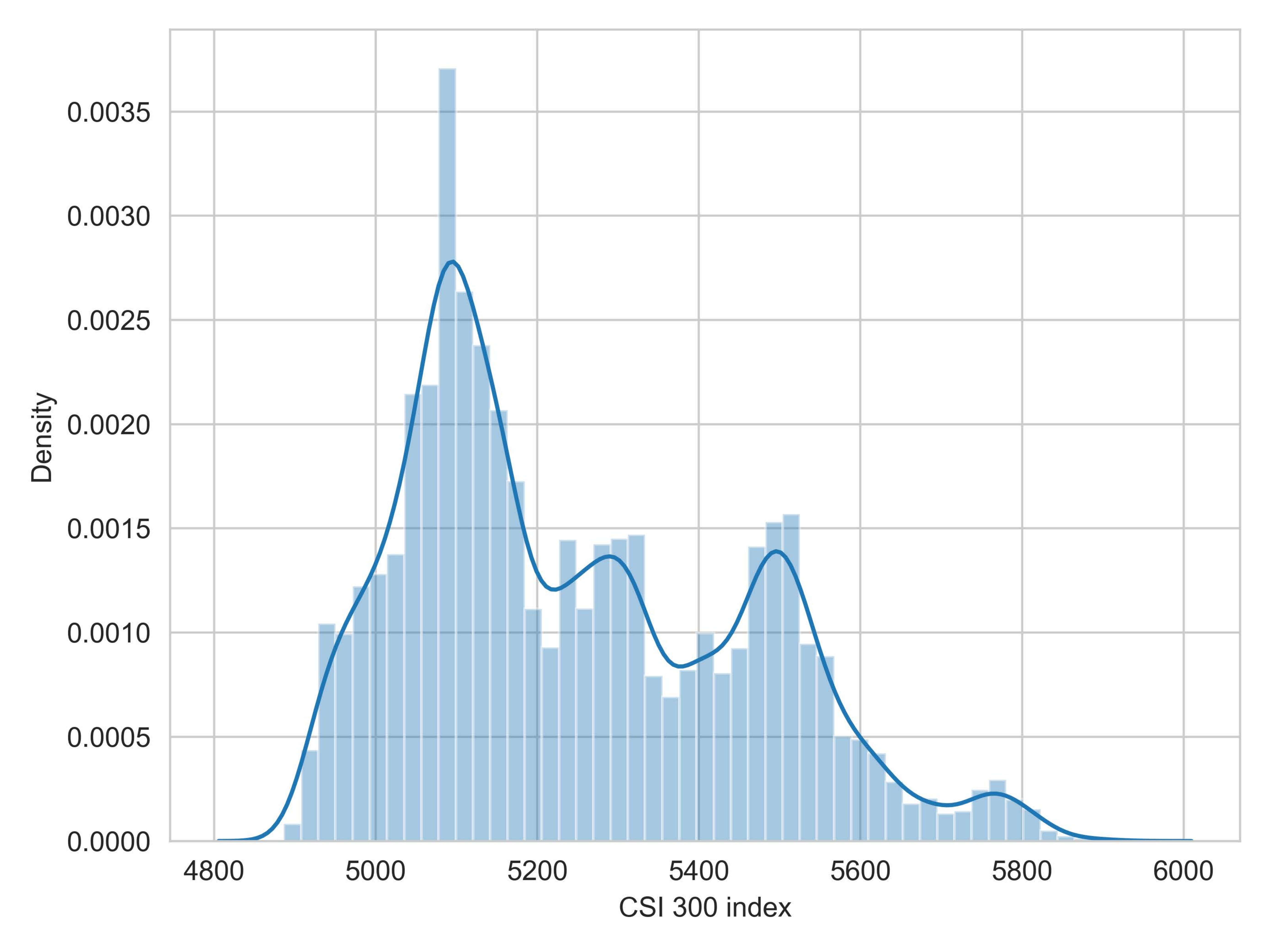}}
\caption{ \small{Distribution plot for $CSI$ $300$ index}}
\end{figure}

Generally, the high-frequency price has a smaller range of changes than the low-frequency price in the same period of time.
For example, the price change per minute is often smaller than the change every two minutes in the same upward trend.
In order to observe small changes in high-frequency data,
the value of the percentage of price change is more suitable to be used as observational data than price data in the analysis of the volatility distribution of high-frequency data.

The histogram of the $CSI$ $300$ price change percentage is shown in Figure 5.
It's seen that $CSI$ $300$ price change statistics per minute, which do not follow the normal distribution,
are mainly concentrated in $0$ (both positive and negative values exist).
The graph is skewed to the right, which indicates that there are more rising empirical data than falling ones in the overall sample.

\begin{figure}[h!]
\centering
{\includegraphics[width=5.5in,height=3.5in]{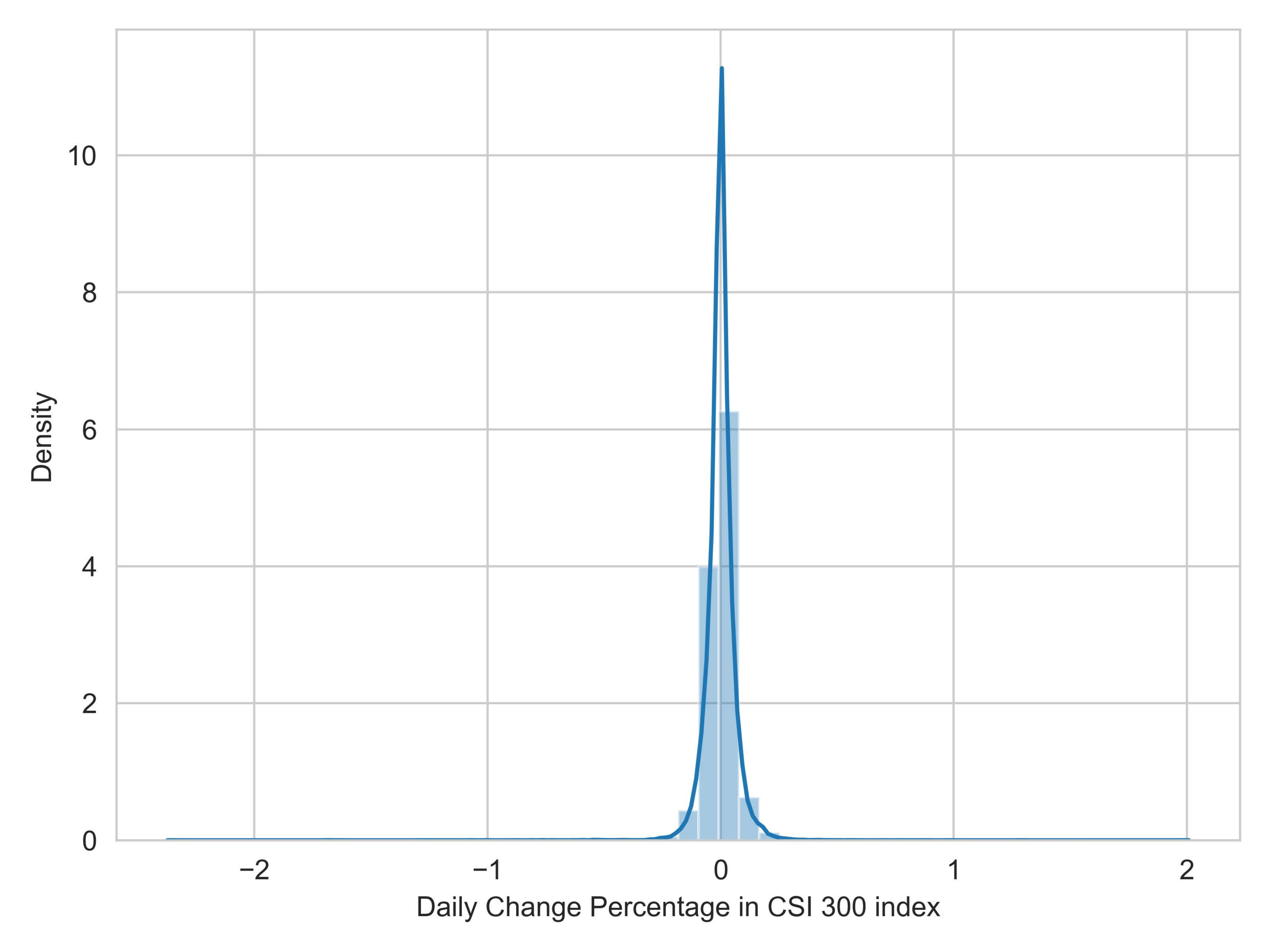}}
\caption{ \small{Histogram for daily change percentage in $CSI$ $300$ index}}
\end{figure}

In order to explore the characteristics of the volatility change of the $CSI$ $300$,
the realized volatility is described separately from the perspective of value distribution and change trend.
Figure 6 is a heatmap of the realized volatility with the sample month as the horizontal axis and the date as the vertical axis.
Through the black and white areas in the figure, the realized volatility date with a volatility change of more than $1\%$ could be identified.
Obviously, the realized volatility of $CSI$ $300$ experienced more frequent fluctuations in each month of the first half of the year.
February contains the largest number of days with large fluctuations.

Figure 7 shows the trend of the realized volatility within half a year. As can be seen from the figure,
the autocorrelation exists in the realized volatility data of the high-frequency $CSI$ $300$ stock index (Volatility Clustering).
The widest range of realized volatility changes occurred in February and March.

Volatility jumps with different amplitudes and frequencies exist in each sample of the $CSI$ $300$.
In the following section,
the information of data characteristics shown in the above charts is used for learning and parameter estimation of empirical data.

\begin{figure}[h!]
\centering
{\includegraphics[width=4.5in,height=3.5in]{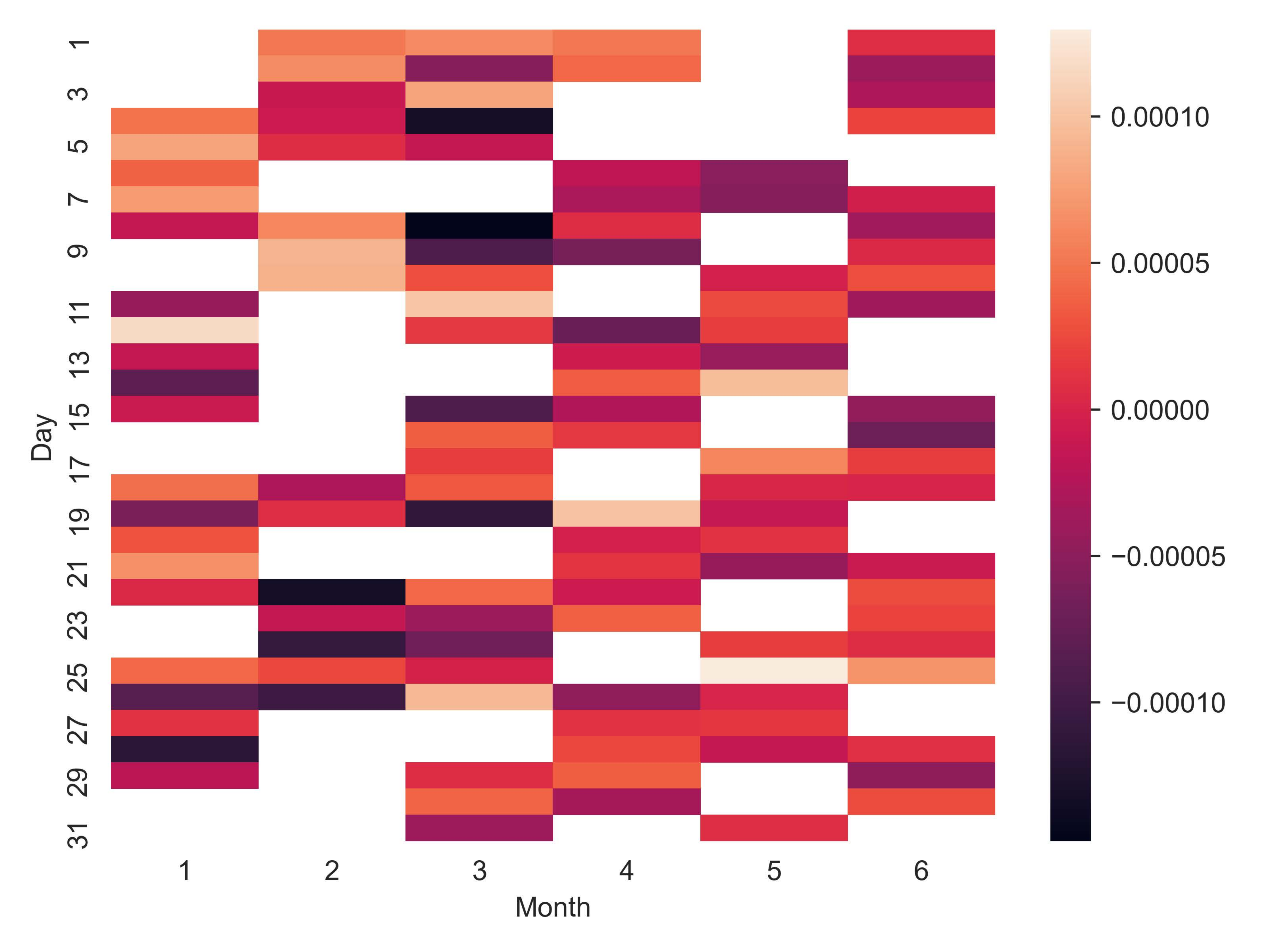}}
\caption{ \small{Heatmap for the realized volatility of high-frequency $CSI$ $300$ index}}
\end{figure}

\begin{figure}[h!]
\centering
{\includegraphics[width=5.5in,height=3in]{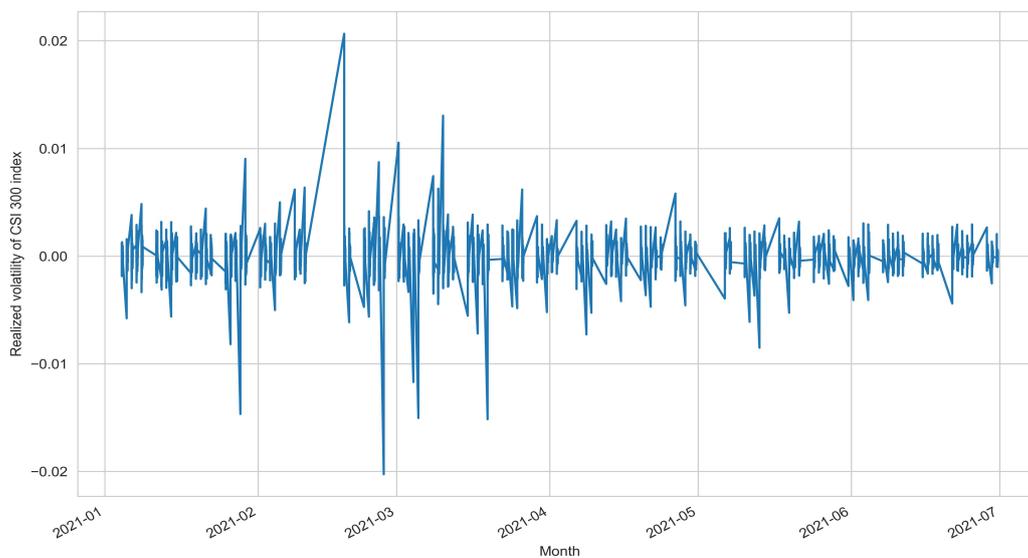}}
\caption{ \small{Line plot for the realized volatility of high-frequency $CSI$ $300$ index}}
\end{figure}

\section{Parameter analysis and estimation}
\label{sec4}
By using the data analysis results in Section 3,
the value of $\theta$ in the generalized new BN-S model in Section 2 is found, and the deterministic component in the random process of high-frequency price data fluctuations is captured in this section.
In order to achieve the above goals, the classification problem based on the historical data set was created by implementing the following steps.

Step 1. Index the available historical price data and price change percentage data per minute of $CSI$ $300$ in chronological order.

Step 2. According to the data fluctuation characteristics obtained in Section 3, create new data structures from historical data sets. Take the percentage of change in the closing price for 10 consecutive minutes as a subset of the rows, stacking layer by layer. Divide the empirical data according to the above rules to form a new $CSI$ $300$ index price data matrix.

Step 3. Consider the volatility of the closing price per minute in the historical price data of the $CSI$ $300$ stock index,
and determine the value of $K$ to define the big jumps (large increases) in the high-frequency closing price fluctuation.
Each time closing price is $K$ lower than the price of the previous minute needs to be identified
(for example, if $K=0.1\%$, the date and time, when the closing price of $CSI$ $300$ index is $0.1\%$ lower than the previous minute's one, should be marked).

Step 4. Create a target column $\theta$ in the new data matrix and assign values. If there are at least two big jumps in the next $10$ minutes, the parameter $\theta$ in the target column of the row is 1. Otherwise, the $\theta$ corresponding to the row is 0.

Step 5. Use several different machine learning algorithms and deep learning algorithms to learn from empirical data sets and estimate the value of $\theta$.
Substitute the obtained value of $\theta$ into (2.8) in Section 2,
which means that the deterministic component of the $CSI$ $300$ fluctuation random process is captured.

The variables involved in the above steps could be adjusted according to the characteristics of the data set.
The adjustment rules could refer to the following reminders.

1. In step 2, if a multi-dimensional data structure is created by adjusting the division of data, the effectiveness of the result will be improved. In general, the more elements contained in a row subset, the more information is carried in the new matrix, and the accuracy of the result may be improved.
At the same time, it also increases the workload of calculation and reduces the predictable time span.

2. In step 3, adjusting the value of $K$ is believed to be an effective way to improve the results.
Different values of $K$ are suitable for different retrieval targets.
Generally, the shorter the time, the smaller the change range of the observed data.
A higher sampling frequency is often suitable for using a smaller value of K, which can identify more big jumps the same period.

3. In step 4, the value of $\theta$ is related to the number of the big jumps identified in a period.

In the same data set, the more subset elements selected in step 2, the more big jumps recognized in each row of the matrix,
and the greater the possibility of $\theta=1$.
Setting the threshold for identifying the number of big jumps to be small will lead to a high probability of $\theta=1$,
and a low probability of $\theta=0$.

The above steps could be used to calculate $\theta$ with reasonable accuracy to prove that these steps are feasible.

Various machine learning and deep learning algorithms are used for the new matrix formed in step 2 on Python. The input are the subset elements in each row of the matrix, and the output is the value of $\theta$ (0 or 1) in the target column of the new matrix mentioned in step 3.
{\color{black} The algorithms we used are $(A)$ Logistic regression, $(B)$ Support vector machine, $(C)$ K-nearest neighbors, $(D)$ K-means, $(E)$ Naive bayes, $(F)$ Gradient boost, $(G)$ Decision tree, $(H)$ Random forest, $(I)$ Neural network, $(J)$ Long and short-term memory neural network (LSTM) and $(K)$ LSTM network with batch normalizer.}
Specifically, logistic regression realize the estimation of $\theta$ through maximum likelihood estimation.
{\color{black} Support vector machine solves the maximum-margin hyperplane of the sample to binary classify the $\theta$ value. K-nearest neighbors predict the classification of $\theta$ by identifying whether there are jumps during the adjacent time intervals. The evaluation of $\theta$ could be realized by calculating the distance of the object and the cluster center in K-means algorithm. Naive bayes estimates the classification of $\theta$ based on bayesian theorem and attribute conditional independence assumption. Gradient boost realize the estimation of $\theta$ by training the new weak classifier on the basis of negative gradient information of the loss function and accumulating the weak classifiers.} After the decision tree is constructed, a reasonable $\theta$ is found through pruning.
Random forest contains many decision trees.
{\color{black} Two hidden and output layers are built in the neural network in algorithm $(I)$.}
If the output probability of the softmax activation function corresponding to $\theta=1$ is greater than $0.3$, the parameter $\theta$ is $1$.
{\color{black} In algorithm $(J)$, Long short-term memory (LSTM) neural network realize forward calculation and back propagation through forward method and backward method.
In algorithm $(K)$, Batch normalizer has a positive effect on the training speed of LSTM.}

Referring to the data characteristics analyzed in Section 3,
four sets of training set and test set dates are selected from the empirical data set to run the classifiers,
and find the corresponding index in step 1. The date selection is shown in Table 2.

\begin{table}[H]
\center
\color{black}
\caption{Time and index of the classifiers}
\vspace{0.5em}
 \setlength{\tabcolsep}{5mm}
 \begin{tabular}{rccc}
 \toprule
& Training Time(Index) 	& Testing Time(Index)\\
 \midrule
T1 & 01/01/2021 9:40:00 (0) to &	02/18/2021 9:40:00 (1316) to\\
 & 02/10/2021 15:00:00 (1315)&	 02/26/2021 15:00:00 (1644)\\
T2 & 01/01/2021 9:40:00 (0) to & 03/01/2021 9:40:00 (1645) to  \\
 & 02/26/2021 15:00:00 (1644)	& 03/12/2021 15:00:00 (2114) \\
T3& 04/01/2021 9:40:00 (2726) to &	06/01/2021 9:40:00 (4559) to \\
 & 05/31/2021 15:00:00 (4558) &	 06/30/2021 15:00:00 (5545) \\
T4& 06/25/2021 9:40:00 (5358) to &	06/29/2021 14:00:00 (5486) to \\
 &  06/29/2021 13:59:00 (5485) &	 06/30/2021 15:00:00 (5545) \\
 \bottomrule
 \end{tabular}
\end{table}

The fluctuation patterns of monthly data, weekly data, daily data, and intraday data are all considered in selecting test samples.
The analysis results in Section 3 show that the average price of sample 2 is the highest, and the volatility is the most intense in February.
Therefore, the daily data of February $(T1)$ where the maximum volatility is located could be selected as the estimated sample.
After experiencing huge ups and downs, $CSI$ $300$ continued to fall in March,
so the price changes in the first two weeks of sample 3 deserve attention and related weekly data $(T2)$ could be estimated.
Monthly data forecast within a more stable range $(T3)$ is also worthy of attention.
It is also meaningful to estimate the parameters on intra-day historical data in the last five days in the data set $(T4)$.
The daily historical data in the last five days in the data set is selected to estimate the value of $\theta$.

In step 5, it is worth noting that the estimation results of $\theta$ by using 6 algorithms are not necessarily the same.
The prediction results of different machine learning and deep learning algorithms often have different accuracy.
In order to avoid possible misjudgments and make the results more accurate,
the prediction results of various algorithms are evaluated.
``Support" refers to the number of responsive samples that appeared during the calculation process.
``Precision" is used to express the accuracy rate in all the prediction results, where $\theta$ takes $1$ or $0$.
It is defined as the ratio of the number of accurate predictions $\theta=1\ (\theta=0)$ to the number of all prediction results $\theta=1 \ (\theta =0)$.
``Recall" shows the efficiency that $\theta=1 \ (\theta =0)$ is accurately predicted.
It represents the ratio of the quantity accurately predicted $\theta=1\ (\theta=0)$ to the true quantity $\theta=1\ (\theta=0)$.
The accuracy of parameter prediction results could be represented by the values of ``precision" and ``recall".
The harmonic average of ``precision" and ``recall" is considered suitable to show the predictive effect of different algorithms directly. Its value is indicated by ``F1-score".

Several machine learning algorithms and deep learning algorithms are used for the empirical data set in the above time periods,
and the results of the classification report of the accuracy evaluation of $\theta$ are recorded in Table 3, Table 4, Table 5, and Table 6.

\begin{table}[H]
\center
\color{black}
\caption{Accuracy report about $\theta$ estimation in $T1$}
\vspace{0.5em}
\resizebox{\textwidth}{30mm}{
\begin{tabular}{ccccccccc} \toprule
 \toprule
 $T1$ & precision & recall & f1-score& support 	& precision & recall	& f1-score	& support 	\\
& $\theta=0$ & $\theta = 0$	& $\theta = 0$	& $\theta = 0$	& $\theta = 1$	& $\theta = 1$	& $\theta = 1$	& $\theta = 1$	\\
 \midrule
(A)& 0.00  &  0.00  & 0.00 &   37&    0.89 &   1.00 &  0.94 &   293 \\
(B)& 0.25  &  0.46  & 0.33  &   37 &    0.92 &  0.83  &  0.87  &  293 \\
(C)& 0.21  & 0.22 &  0.21  &  37 &   0.90  &  0.90  & 0.90  &  293 \\
(D)& 0.16  & 0.11  &  0.13   &  37  &   0.89  & 0.11   & 0.20   & 293 \\
(E)& 0.03  & 0.03  & 0.03   & 37  &  0.88   & 0.87   & 0.88   & 293 \\
(F)& 0.17  & 0.03 &  0.05    & 37   &  0.89 &   0.98 &   0.93 &   293 \\
(G)& 0.15  &  0.41 &  0.22 &   37   & 0.90 &  0.71 &  0.80 &   293 \\
(H)& 0.14  & 0.05 &  0.08  &   37    & 0.89  &  0.96  &  0.92  &  293 \\
(I)& 0.17  &  0.05  &  0.08  &  37  &   0.89  & 0.97  & 0.93  &  293 \\
(J)& 0.00  &  0.00  & 0.00   &  37  &  0.89   & 0.99  &  0.93 &  293 \\
(K)& 0.14  & 0.03  & 0.05   & 37   &  0.89   &0.98  & 0.93  &  293 \\
\bottomrule
\end{tabular}}
\end{table}

\begin{table}[H]
\center
\color{black}
\caption{Accuracy report about $\theta$ estimation in $T2$}
\vspace{0.5em}
\resizebox{\textwidth}{30mm}{
\begin{tabular}{ccccccccc} \toprule
 \toprule
 $T2$ & precision & recall & f1-score& support 	& precision & recall	& f1-score	& support 	\\
& $\theta=0$ & $\theta = 0$	& $\theta = 0$	& $\theta = 0$	& $\theta = 1$	& $\theta = 1$	& $\theta = 1$	& $\theta = 1$	\\
 \midrule
(A)	& 0.00 	&   0.00 	&   0.00 	&    72  	&   0.85  	&  1.00  	&  0.92  	&   399 \\
(B)	& 0.08 	&   0.10  	&  0.09  	&   72   	&  0.83 	&   0.79 	&   0.81  	&  399 \\
(C)	& 0.21  	&  0.29  	&  0.25  	&   72   	&  0.86  	&  0.80 	&   0.83  	&  399 \\
(D) 	& 0.19  	&  0.11   	& 0.14   	&  72  	&   0.92  	&  0.12 	&   0.22  	&  399 \\
(E) 	& 0.36  	&  0.68  	&  0.47  	&   72  	&   0.93 	&   0.78  	&  0.85  	&   399 \\
(F)	& 0.36 	&   0.11  	&  0.17   	&  72  	&   0.86  	&  0.96  	&  0.91  	&   399 \\
(G)	& 0.17  	&  0.33 	&   0.22  	&   72  	&   0.85  	&  0.70  	&  0.77  	&  399 \\
(H)	& 0.23  	&  0.08  	&  0.12   	&  72  	&   0.85  	&  0.95  	&  0.90 	&   399 \\
(I)	& 0.18 	&   0.08  	&  0.11  	&   72  	&    0.85 	&   0.93  	&  0.89  	&   399 \\
(J)	& 0.26  	&  0.17  	&  0.20   	&  72   	&   0.86 	&   0.91  	&  0.88  	&   399 \\
(K)	& 0.36  	&  0.25  	&  0.30 	&    72   	&   0.87  	&  0.92 	&   0.90  	&   399 \\
\bottomrule
\end{tabular}}
\end{table}

\begin{table}[H]
\center
\color{black}
\caption{Accuracy report about $\theta$ estimation in $T3$}
\vspace{0.5em}
\resizebox{\textwidth}{30mm}{
\begin{tabular}{ccccccccc} \toprule
 \toprule
 $T3$ & precision & recall & f1-score& support 	& precision & recall	& f1-score	& support 	\\
& $\theta=0$ & $\theta = 0$	& $\theta = 0$	& $\theta = 0$	& $\theta = 1$	& $\theta = 1$	& $\theta = 1$	& $\theta = 1$	\\
 \midrule
(A)& 0.42 &  0.34 &  0.38 &   454  &  0.51  &  0.59 &  0.54  &  525 \\
(B) &0.45 &  0.51 &  0.48 &   454  &  0.52  &  0.46 &  0.49  &  525 \\
(C)& 0.48 &  0.58 &  0.53 &   454  &  0.56  &  0.45 &  0.50  &  525 \\
(D)& 0.43 &  0.09 &  0.15 &   454  &  0.48  &  0.14 &  0.21  &  525 \\
(E)& 0.54 &  0.78 &  0.64 &   454  &  0.69  &  0.43 &  0.53  &  525 \\
(F)& 0.49 &  0.53 &  0.51 &   454  &  0.57  &  0.53 &  0.55  &  525 \\
(G)& 0.48 &  0.49 &  0.49 &   454  &  0.55  &  0.54 &  0.55  &  525 \\
(H)& 0.49 &  0.57 &  0.53 &   454  &  0.57  &  0.50 &  0.53  &  525 \\
(I)& 0.46 &  0.31 &  0.37 &   454  &   0.53 &  0.68 &   0.60 &   525 \\
(J)& 0.47 &  0.34 &  0.40 &   454  &   0.54 &  0.67 &   0.60 &   525 \\
(K)& 0.45 &  0.24 &  0.31 &   454  &   0.53 &  0.74 &   0.62 &   525 \\
\bottomrule
\end{tabular}}
\end{table}

\begin{table}[H]
\center
\color{black}
\caption{Accuracy report about $\theta$ estimation in $T4$}
\vspace{0.5em}
\resizebox{\textwidth}{30mm}{
\begin{tabular}{ccccccccc} \toprule
 \toprule
 $T4$ & precision & recall & f1-score& support 	& precision & recall	& f1-score	& support 	\\
& $\theta=0$ & $\theta = 0$	& $\theta = 0$	& $\theta = 0$	& $\theta = 1$	& $\theta = 1$	& $\theta = 1$	& $\theta = 1$	\\
 \midrule
(A)& 0.78  &  0.90  &  0.83  &   39  &  0.43  &  0.23  &  0.30  &  13 \\
(B)& 0.77  &  0.92  &  0.84  &   39  &  0.40  &  0.15  &  0.22  &  13 \\
(C)& 0.76  &  0.90  &  0.82  &   39  &  0.33  &  0.15  &  0.21  &  13 \\
(D)& 0.80  &  0.10  &  0.18  &   39  &  0.09  &  0.08  &  0.08  &  13 \\
(E)& 0.83  &  0.90  &  0.86  &   39  &  0.60  &  0.46  &  0.52  &  13 \\
(F)& 0.70  &  0.67  &  0.68  &   39  &  0.13  &  0.15  &  0.14  &  13 \\
(G)& 0.74  &  0.59  &  0.66  &   39  &  0.24  &  0.38  &  0.29  &  13 \\
(H)& 0.79  &  0.85  &  0.81  &   39  &  0.40  &  0.31  &  0.35  &  13 \\
(I)& 0.78  &  0.72  &  0.75  &   39  &   0.31 &  0.38 &  0.34  &   13 \\
(J)& 0.81  &  0.64  &  0.71  &   39  &   0.33 &  0.54 &  0.41  &   13 \\
(K)& 0.84  &  0.69  &  0.76  &   39  &   0.40 &  0.62 &  0.48  &   13  \\
\bottomrule
\end{tabular}}
\end{table}

It can be seen that the number of ``support" in the report results, not affected by different algorithms, is only related to the time window T. It shows the number of jumps could be accurately identified by the {\color{black}eleven algorithms we used for the empirical price data of $CSI 300$ Index.} Comparing the number of ``supports" (T4 \textless T1 \textless T2 \textless T3), we find that as the time window T grows, the more jumps would be identified. It's just like we thought.

The classification results in Table 3 illustrate that when threshold of the asset return K for identifying jumps is 0.1, {\color{black}there’s a strong possibility that $\theta$ is equal to 1, while $\theta=0$ is still a possibility,} not a probability. And the results in Table 4, Table 5 and Table 6 have reached the same conclusion.

The difference, however, is that the probability of $\theta=1$ is different when using different algorithms in different time windows. Comparing the results in Table 3 and Table 5, the possibility of $\theta=1$ in the parameter estimation results of daily data is greater than that of monthly data. It shows that the price fluctuation in T1 is wider than that in T3, which is also entirely consistent with our analysis results in Section 3. The similar conclusions can be obtained in the comparison of Table 4 and Table 6.

Machine learning algorithms can more effectively extract information from large amounts of structured or unstructured data, and analyze data with less understanding of data structure or the relationship between input and output (including nonlinear relationship) to realize the quantification and prediction of high-frequency transaction data. Machine learning and deep learning algorithms automate the decision-making process to overcome the limitations of human decision-making. Machine learning and deep learning algorithms automate the decision-making process to overcome the limitations of human decision-making. The efficiency of the analysis results shows the advantages of using machine learning and deep learning algorithms to deal with nonlinear and collinear relations, the accuracy of stock price fitting and the ability to predict future jumps.

In the end, the results of the classification reports could be used to determine the value of $\theta$. After the dynamic value of $\theta$, as deterministic component in stochastic process of $CSI$ $300$ index price fluctuations, {\color{black} is substituted into formula \eqref{(2.10)}, the dynamic process of the price fluctuation of $CSI$ $300$ index would be described flexibly and effectively by the generalized BN-S model, formulas \eqref{(2.9)}, \eqref{(2.10)}, \eqref{(2.11)}.}

\section{Conclusions}
\label{sec5}
The fluctuation of price is a regular phenomenon, which has been empirically seen to widely exist in financial markets.
{\color{black} This paper introduces a new generalized BN-S model to describe stochastic fluctuations in high-frequency asset price dynamics.}
The new model considers the lag caused by the asynchrony of market information, and adds new parameters to the classic BN-S model, which effectively expands the application range.

The high-frequency $CSI$ $300$ index in Chinese financial market was selected as the research sample.
The empirical data was preprocessed (the influence of overnight information on the market was removed),
and a series of statistical analyses were performed to estimate its volatility characteristics.
With the help of machine learning and deep learning algorithms, we analyzed dynamic prices in different time spans (monthly data, weekly data, daily data and intraday data), and estimated the deterministic component in the stochastic fluctuation process of high-frequency price data to show good operability of the new model.

Our work provides a new perspective for the analysis of price fluctuations in the financial sector,
and is of positive significance for improving the accuracy of dynamic fluctuation estimation.
In ongoing work, we are working on extensions of new models to accommodate to more complicated financial market scenarios and investor needs.
Future research will include, but not be limited to, the study of predicting the existence and magnitude of dynamic jumps in high-frequency fluctuations, confirming quantified trading timing and avoiding the risk of abnormal fluctuations.

\section*{Acknowledgments}
This work is jointly supported
by the National Natural Science Foundation of China under Nos. 11662001 and 11771105, the Science Foundation of Guangxi Province under
Nos. 2017GXNSFFA198012 and 2018GXNSFAA138177.

\bibliographystyle{siam}
\bibliography{mycite}

\end{document}